\magnification=1000 \baselineskip=12pt plus 1pt minus 2pt
%%%%%%%%%%%%%%%%%%%%%%%%%%%%%%%%%%%%%%%%%%%%
\def\sqr#1#2{{\vcenter{\vbox{\hrule height.#2pt
       \hbox{\vrule width.#2pt height#1pt \kern#1pt
          \vrule width.#2pt}
        \hrule height.#2pt}}}}
\def\square{\mathchoice\sqr55\sqr55\sqr{2.1}3\sqr{1.5}3}
%%%%%%%%%%%%%%%%%%%%%%%%%%%%%%%%%%%%%%%%%%%%

\parindent=2.0 truecm
\centerline{\bf BRST QUANTIZATION}

\centerline{\bf OF THE MASSLESS MINIMALLY COUPLED SCALAR FIELD IN DE
SITTER SPACE}

\centerline{\bf (ZERO MODES, EUCLIDEANIZATION  AND QUANTIZATION)}

\bigskip
\bigskip
\centerline{Antoine Folacci}
\smallskip
\centerline{Universit\'e de Corse, Facult\'e des Sciences, B.P.52,}
\centerline{20250 Corte, France}
\bigskip
\vskip 2.0truecm

\noindent NOTE: This paper has been published under the title ``Zero
modes, euclideanization and quantization" [Phys. Rev. D46, 2553
(1992)].

\bigskip\bigskip

\noindent ABSTRACT: We consider the massless scalar field on the
four-dimensional sphere $S^4$. Its classical action $S={1\over
2}\int _{S^4} dV~(\nabla \phi )^2$ is degenerate under the global
invariance $\phi \rightarrow \phi + \hbox{constant}$. We then
quantize the massless scalar field as a gauge theory by constructing
a BRST-invariant quantum action. The corresponding gauge-breaking
term is a non-local one of the form $S^{\rm GB}={1\over {2\alpha
V}}\bigl(\int _{S^4} dV~\phi \bigr)^2$ where $\alpha$ is a gauge
parameter and $V$ is the volume of $S^4$. It allows us to correctly
treat the zero mode problem. The quantum theory is invariant under
$SO(5)$, the symmetry group of $S^4$, and the associated two-point
functions have no infrared divergence. The well-known infrared
divergence which appears by taking the massless limit of the massive
scalar field propagator is therefore a gauge artifact. By contrast,
the massless scalar field theory on de Sitter space $dS^4$ - the
lorentzian version of $S^4$ - is not invariant under the symmetry
group of that spacetime $SO(1,4)$. Here, the infrared divergence is
real. Therefore, the massless scalar quantum field theories on $S^4$
and $dS^4$ cannot be linked by analytic continuation. In this case,
because of zero modes, the euclidean approach to quantum field
theory does not work. Similar considerations also apply to massive
scalar field theories for exceptional values of the mass parameter
(corresponding to the discrete series of the de Sitter group).

\bigskip

\vfill
\eject

\baselineskip=15pt plus 1pt minus 2pt
\parindent=1.5 truecm

\centerline {\bf 1 - INTRODUCTION}
\bigskip

          The euclidean approach (i.e.~generalization of Wick rotation) to
quantum field theory in curved spacetimes has been extensively used
in particular i) in order to get Feynman propagators or anticommutator functions
in an elegant way, ii) in connection with path integral quantization, and iii)
in the context of quantum cosmology. (See [1,2,3,4,5] and references therein for
more details.) Its principal advantages are the following: it permits one to deal
with elliptic operators instead of hyperbolic ones and then to consider
well-posed problems and mathematically well-defined objects and expansions. It
also permits one to consider path integrals with well-defined measure on the
space of paths and which are convergent rather than oscillating and divergent. In
simple cases such as de Sitter, Anti-de Sitter or Schwarzschild spacetimes as
well as globally static ones (and more generally for certain spacetimes which
can be considered as sections of four-dimensional complex manifolds), it seems
that the euclidean approach does not present any difficulties, at least if
``boundary conditions" are considered with care.

\medskip

          In this paper, we study the quantization of the massless minimally
coupled scalar field on the euclidean version $S^4$ of de Sitter
space $dS^4$. We BRST-quantize that theory. It is characterized by a
non-local gauge-breaking term. The corresponding ghost field $c$,
antighost field ${\overline c}$ and auxillary field $b$ are all
constant. In path integrals, the non-propagating auxillary field
must be integrated in a complex direction (as the conformal factor
of gravitation [2]). We calculate the two-point function $\langle
\phi (x) \phi (x') \rangle $ associated with the scalar field. We
show that its well-known infrared divergence is only a gauge
artifact and that it exhibits an $SO(5)$-symmetry. ($SO(5)$ is the
symmetry group of $S^4$.) We evaluate the associated renormalized
stress-energy tensor; it is noted that the contributions of the
fields $c$,${\overline c}$ and $b$ cancel. The physical theory
obtained by working on the ordinary version of de Sitter space does
not possess a similar symmetry. Because of zero modes, a
$SO(1,4)$-invariant propagator necessarily presents an infrared
divergence [6]. (Let us recall that in that case, the breakdown of
$SO(1,4)$-symmetry and the time-dependence of quantities such as
$\langle \phi ^2 \rangle $ has a great importance in the context of
the cosmological inflation.) Thus, that physical theory cannot be
obtained by analytic continuation from its euclidean counterpart. In
that case, the euclidean approach cannot be used to understand the
physical theory.

\medskip

        In an appendix, we extend (on $S^4$) the results obtained for the
massless scalar field theory: we BRST-quantize the massive scalar field theories
corresponding to the exceptional values ${m_p}^2=-p(p+3)(R/12)$ of the mass
parameter. (Here $R$ is the scalar curvature of $S^4$ and $p \in
{\bf N^*}$.) All these theories present finite-dimensional gauge invariances.

\vfill
\eject

\centerline {\bf 2 - QUANTIZATION OF THE MASSLESS SCALAR FIELD ON $S^4$}
\bigskip

          In order to understand the so-called infrared divergence which
appears in the massless minimally coupled scalar field theory in de Sitter
space, let us first consider the massive scalar field. Its euclidean action is
given by
$$S(\phi)={1\over 2}\int _{S^4} dV~\bigl[ (\nabla \phi
)^2+m^2\phi ^2\bigr]={1\over 2}\int _{S^4} dV~\bigl( -\phi \square
\phi +m^2\phi ^2\bigr), \eqno({\rm 2}.1)$$ \noindent where
$dV=(g)^{1/2}d^4x$. The four-dimensional sphere is characterized by
a radius ${1/H}$ and therefore by a scalar curvature $R=12H^2$. Its
volume is given by
$$V=\int _{S^4} dV={8\pi ^2  \over  3H^4 }.\eqno({\rm 2}.2) $$
\noindent In order to evaluate path integrals over $\phi$, we shall
decompose $\phi$ on the complete set of the eigunfunctions of the
laplacian $\square$. Because $S^4$ is a compact Riemannian manifold,
$\square$ possesses a discrete spectrum of eigenvalues $\lambda _n$.
The corresponding eigenfunctions $\phi _n^i$ (see for example [7])
are such that
$$\eqalignno{&\square \phi _n^i =-\lambda _n \phi _n^i \qquad ~\quad i=1,...,d_n
&({\rm 2}.3)\cr &\lambda _n = H^2n(n+3) \qquad n=0,1,2,... &({\rm
2}.4)\cr}$$
\noindent The degeneracy of each eigenvalue $\lambda _n$
is $d_n={1\over 6}(n+1)(n+2)(2n+3)$. Moreover, without loss of
generality, the $\phi _n^i$ may be taken real and orthonormalized.
We then have
$$\eqalignno{&\int _{S^4} dV~ \phi _n^i \phi _m^j=\delta _{nm}\delta _{ij},
&({\rm 2}.5)\cr &\sum _n \sum _{i=1}^{d_n} \phi _n^i (x) \phi _n^i
(x')= \delta ^4 (x,x'). &({\rm 2}.6)\cr}$$
\noindent Now, in order
to simplify our notation, we will suppress all degeneracy indices,
but in the following, all the sums and products over $n$ must be
understood as sums and products over $n$ and $i$. (In the appendix,
it will be necessary to reintroduce the degeneracy indicies.) It
should be noted that the lowest eigenvalue of the laplacian
$\square$ is $\lambda _0=0$. Its unique associated eigenfunction
$\phi _0$ (zero mode) is a constant given by the normalization
relation (2.5):
$$\phi _0=V^{-1/2}={\sqrt {3\over 8\pi ^2}}H^2. \eqno({\rm 2}.7)$$
\noindent Moreover, in the following, we shall also use the relation
$$\int _{S^4} dV~ \phi _n =0 \quad \hbox{if}~n\not= 0 \eqno({\rm 2}.8)$$
\noindent which is a direct consequence of (2.6). By expanding the
field $\phi$ on the complete set of the $\phi _n$ eigenfunctions as
$$\phi = \sum _n a_n \phi _n  \eqno({\rm 2}.9)$$
\noindent and by using the relations (2.3) and (2.5) we get for the
action (2.1)
$$S(\phi)={1\over 2}\sum _n (\lambda _n + m^2){a_n}^2. \eqno({\rm 2}.10)$$
\noindent The two-point function $G(x,x';m^2)=\langle \phi (x) \phi
(x') \rangle$ is obtained from
$$G(x,x';m^2)={{\int d[\phi ]~ \phi (x) \phi (x') \exp (-S)}\over {\int d[\phi
]~\exp (-S)}} \eqno({\rm 2}.11)$$
\noindent where the measure
$d[\phi ]$ on the space of fields is
$$d[\phi ]=\prod _n da_n .\eqno({\rm 2}.12)$$
\noindent By inserting (2.9) and (2.10) into (2.11) and by using the
relations
$$\int _{-\infty}^{+\infty} dx ~\exp (-\alpha x^2)=\sqrt {{\pi \over
\alpha}}, \quad  \int _{-\infty}^{+\infty} dx ~x\exp (-\alpha x^2)=0
\quad \hbox{and} \quad  \int _{-\infty}^{+\infty} dx ~x^2\exp
(-\alpha x^2)=~{1\over {2\alpha}}\sqrt {{\pi \over \alpha}},$$
\noindent one finds that
$$G(x,x';m^2)=\sum _n {{\phi _n(x) \phi _n(x')}ÊÊ\over {\lambda _n + m^2} }.
\eqno({\rm 2}.13)$$
\noindent It is possible to perform the sum in
(2.13). One then finds the usual result [8] giving the euclidean
two-point function $G(x,x';m^2)$:
$$G(x,x';m^2)={R\over 192\pi ^2}\Gamma ({3/2}+\nu)\Gamma ({3/2}-\nu)F[
{3/2}+\nu,{3/2}-\nu;2;Z(x,x')]\eqno({\rm 2}.14)$$ \noindent where
$\nu ^2={9\over 4}-{m^2 \over H^2}$ and $Z(x,x')=\cos ^2\bigl[H\mu
(x,x')/2 \bigl]$. Here $\mu (x,x')$ is the geodesic distance between
the points $x$ and $x'$ on $S^4$. Because $\mu (x,x')$ and $Z(x,x')$
are invariant under the symmetry group $SO(5)$, $G(x,x';m^2)$ is
also $SO(5)$-invariant.

\medskip

          In the massless limit, $G(x,x';m^2)$ given by (2.14) is divergent. From [9], we obtain
$$G(x,x';m^2)={R^2\over 384\pi ^2m^2}
+{R\over 96\pi ^2}\biggl[{{1/2}\over 1-Z(x,x')}- \ln \bigl(1-
Z(x,x')\bigr) \biggr] + {\cal O}(m^2).\eqno({\rm 2}.15)$$ \noindent
It is obvious that this infrared divergence comes from the zero mode
$\phi _0$. Indeed, by considering the massless limit of (2.13), we
see that all the denominators in that expression are nonzero except
those associated with $n=0$. Therefore the expression $\sum _{n\not=
0} {{\phi _n(x) \phi _n(x')}Ê\over {\lambda _n + m^2} }$ is finite
while ${{\phi _0(x) \phi _0(x')}Ê\over {\lambda _0 + m^2} }$
diverges like ${1\over m^2}$ because $\lambda _0 =0$. It should be
noted that from the expansion (2.15) and the relation (2.7) we get
$$\sum _{n\not= 0} {{\phi _n(x) \phi _n(x')}ÊÊ\over {\lambda _n}}={R\over 96\pi
^2}\biggl[{{1/2}\over1- Z(x,x')}- \ln \bigl(1- Z(x,x')\bigr)
\biggr].  \eqno({\rm 2}.16)$$
\noindent The infrared divergence is
also present in the partition function ${\cal Z}=\int d[\phi]\exp
(-S)$: when $m^2=0$, the partition function ${\cal Z}=\int
d[\phi]\exp (-S)$ diverge because in its expression the term $ \int
_{-\infty}^{+\infty} da_0\exp [-{1\over 2}(\lambda _0 +
m^2){a_0}^2]$ reduces to $ \int _{-\infty}^{+\infty} da_0$.

\medskip

          Now, we shall prove that this infrared divergence is a gauge artifact.
For $m^2=0$, the action (2.1) becomes
$$S(\phi)={1\over 2}\int _{S^4} dV~ (\nabla \phi
)^2=-{1\over 2}\int _{S^4} dV~\phi \square \phi  \eqno({\rm 2}.17)$$
\noindent and is invariant under the one-dimensional gauge symmetry
$\phi \rightarrow \phi + \hbox{constant}$. The quantization of that
theory can be realized by using BRST methods in the spirit of
[10,11,12]. We consider a fermionic operator ${\bf s}$ constructed
such that ${\bf s}^2=0$ and defined by its action on the field
$\phi$ and on all the algebraic fields introduced at the quantum
level: we have
$${\bf s}\phi =c, ~~{\bf s}c =0, ~~{\bf s}{\overline c}=b, ~~{\bf s}b =0.
\eqno({\rm 2}.18)$$
\noindent Here $c$ is the anticommutating ghost
field associated to the invariance $\phi \rightarrow \phi +
\hbox{constant}$ and it is constant. ${\overline c}$ is a constant
anticommutating antighost and $b$ is a constant commuting auxillary
field. The relation ${\bf s}c =0$ arises on this simple form because
the gauge transformation is an abelian one while the relations ${\bf
s}{\overline c}=b$ and ${\bf s}b =0$ are usual in the BRST
formalism. The operator ${\bf s}$ must be interpreted as a linear
differential operator graded by the ghost number. (The ghost number
is 0 for $\phi$ and $b$, $+1$ for $c$ and $-1$ for ${\overline c}$.
The ghost number of a product of fields is the sum of the ghost
numbers of the fields.) We then have ${\bf s}(AB)={\bf s}(A)B+
(-1)^{n(A)}A{\bf s}(B)$ where $n(A)$ denotes the ghost number of
$A$. Moreover, we suppose that ${\bf s}$ commutes with spacetime
variables and spacetime derivatives. In order to quantize the
massless scalar field theory, we add to the classical action (2.17)
the following gauge-fixing term
$$S^{GF}={\bf s}\int _{S^4} dV~\bigl({\overline c}\phi - {1\over 2}\alpha {\overline
c}b \bigr)\eqno({\rm 2}.19)$$
\noindent where $\alpha$ is a gauge
parameter. $S^{GF}$ is ${\bf s}$-exact and therefore ${\bf
s}$-invariant. Moreover, the classical action (2.17) is also ${\bf
s}$-invariant and thus the total quantum action is ${\bf
s}$-invariant. Furthermore, the fact that in the total action the
gauge parameter $\alpha$ is the coefficient of a ${\bf s}$-exact
term which is analytic in the fields ensures the gauge independence
of the quantum theory (at least in the tree approximation) [10,11].
We obtain a better interpretation of $S^{GF}$ by using (2.18) to
obtain
$$S^{GF}(\phi,c,{\overline c},b)=\int _{S^4} dV~\bigl(b\phi -{\overline c}c - {1\over 2}\alpha
b^2 \bigr)=b\biggl(\int _{S^4} dV~\phi \biggr) - V{\overline c}c -
{1\over 2}\alpha V b^2 \eqno({\rm 2}.20)$$
\noindent and then by
performing the shift $b\rightarrow b + {1\over \alpha V}\int _{S^4}
dV~\phi$. (It should be noted that the Jacobian of the variable
change $(\phi ,c,{\overline c},b) \rightarrow (\phi ,c,{\overline
c},b+{1\over \alpha V}\int _{S^4} dV~\phi)$ is equal to one. Thus,
the measure on the space of all the fields $d[\phi]~d{\overline
c}~dc~db$ remains inchanged.) The total quantum action becomes
$$S^Q(\phi,c,{\overline c},b)={1\over 2}\int _{S^4} dV~ (\nabla \phi
)^2 + {1\over {2\alpha V}}\biggl(\int _{S^4} dV~\phi \biggr)^2
-V{\overline c}c - {1\over 2}\alpha V b^2.\eqno({\rm 2}.21)$$
\noindent The second term in the right-hand side of (2.21) clearly
appears as a non-local gauge-breaking term. It breaks the invariance
$\phi \rightarrow \phi + \hbox{constant}$.

\medskip

          In the following, we shall evaluate functional integrals by summing
over $\phi$, ${\overline c}$, $c$ and $b$ (with this order) expressions of
the form $f(\phi ,c,{\overline c},b)\exp (-S^Q)$. In order to get convergent
integrals over $\phi$, the gauge parameter $\alpha$ has to be taken positive.
But then the last term in (2.21) is problematic. If the integration over $b$ is
taken on the real axis, the path integral diverges. In order to get convergent
path integrals, it is necessary to adopt the following prescription: the
integration over $b$ has to be taken in the imaginary complex direction. In an
equivalent way, we must change the sign in front of the last term of (2.21). A
similar problem exists in the path integral approach of quantum gravity [2]. In
that case, in order to get convergent integrals, the integration over the
conformal factor has to be taken also in a complex direction. It is important to
understand that the proposed prescription is not an artificial way to eliminate
the infrared divergence. The infrared divergence problem and the problem of the
divergence of the integrals over $b$ are totally different. We believe that the
second problem arises because of the nature of the auxillary field: like the
conformal factor of gravitation, it is not a propagating field. Exactly for the
same reasons, we change the sign in front of the ghost term in the
action. (See [13] for a complementary discussion on the integration over $c$,
${\overline c}$ and $b$.) In conclusion, in the path integrals we shall consider
the positive definite action
$$S^Q(\phi,c,{\overline c},b)={1\over 2}\int _{S^4} dV~ (\nabla \phi
)^2 + {1\over {2\alpha V}}\biggl(\int _{S^4} dV~\phi \biggr)^2 +
V{\overline c}c + {1\over 2}\alpha V b^2 \eqno({\rm 2}.22)$$
\noindent and we shall integrate over the real values of $\phi$ and
$b$, and over the grassmannian variables ${\overline c}$ and $c$
(with this order) by using the usual rules $\int dc =0$, $\int
d{\overline c} =0$, $\int dc~c =1$ and $\int d{\overline
c}~{\overline c} =1$.

\medskip

        Let us first consider the partition function ${\cal Z}$ of the massless scalar
field theory. It is defined as
$${\cal Z}=\int d[\phi]~d{\overline c}~dc~db~\exp(-S^Q).\eqno({\rm 2}.23)$$
\noindent By writing $\phi = \sum _n a_n \phi _n$ and from (2.3),
(2.5), (2.7) and (2.8), it is obvious that
$$S^Q(\phi,c,{\overline c},b)={1\over 2}\sum _{n\not= 0} \lambda _n {a_n}^2 +
{1\over {2\alpha }}{a_0}^2 + V{\overline c}c + {1\over 2}\alpha V
b^2 \eqno({\rm 2}.24)$$ \noindent and therefore we get
$${\cal Z}=2\pi V^{1/2}\prod _{n\not= 0} \biggl({2\pi \over \lambda _n}
\biggr)^{1/2}. \eqno({\rm 2}.25)$$ \noindent ${\cal Z}$ is
independent of the gauge parameter $\alpha$. Moreover, it is not
infrared divergent. It needs only a regularization because of the
usual ultraviolet divergence of the term $\prod _{n\not= 0}
\bigl({2\pi \over \lambda _n}\bigr)^{1/2}$.

\medskip

           The two-point function $G(x,x')=\langle \phi (x) \phi (x')
\rangle$ is now given by
$$G(x,x')={{\int d[\phi ]~d{\overline c}~dc~db~ \phi (x) \phi (x') \exp
(-S^Q)}\over {\int d[\phi ]~d{\overline c}~dc~db~\exp (-S^Q)}}.
\eqno({\rm 2}.26)$$
\noindent From (2.24) and by inserting $\phi =
\sum _n a_n \phi _n$ in (2.26) we obtain
$$G(x,x')=\sum _{n\not= 0} {{\phi _n(x) \phi _n(x')}ÊÊ\over {\lambda _n} } +
\alpha {\phi _0(x) \phi _0(x')} \eqno({\rm 2}.27)$$
\noindent and
from (2.7) and (2.16) one finds
$$G(x,x')={R\over 96\pi
^2}\biggl[{{1/2}\over 1-Z(x,x')}- \ln \bigl(1-Z(x,x')\bigr)  \biggr]
+ \alpha \biggl({3H^4 \over 8\pi ^2}\biggr).  \eqno({\rm 2}.28)$$
\noindent Clearly, $G(x,x')$ is finite and is $SO(5)$-invariant. The
so-called infrared divergence is nothing but a gauge artefact. It
occurs when the gauge parameter $\alpha$ goes to $\infty$.
Similarily, the Feynman propagator and the anticommutator function
are also finite and $SO(5)$-invariant. Moreover, these two last
Green functions possess Hadamard expansions. It should be noted that
all these two-point functions depend on the gauge parameter $\alpha$
but that the physical quantities calculated from them must be gauge
parameter independent (even in the limit $\alpha \rightarrow
\infty$). The choice (2.20) for the gauge-fixing action ensures it.

\medskip

          An important example of a physical quantity is provided
by $\langle T_{\mu \nu} \rangle$, the vacuum expectation value of the
stress-energy tensor. The stress-energy operator is formally constructed from
$S^Q$ by
$$T_{\mu \nu}=-{2\over {g^{1/2}}}{\delta S^Q \over \delta g^{\mu \nu}}
\eqno({\rm 2}.29)$$
\noindent and by using the fact that in the
transformation $g_{\mu \nu} \rightarrow g_{\mu \nu} + \delta g_{\mu
\nu}$ we have $g^{1/2} \rightarrow g^{1/2} + {1\over 2}g^{1/2}g_{\mu
\nu}\delta g_{\mu \nu}$, we get
$$T_{\mu \nu}=T_{\mu \nu}^{\rm Cl} + T_{\mu \nu}^{\rm GB} + T_{\mu
\nu}^{\rm g} +T_{\mu \nu}^{\rm b}\eqno({\rm 2}.30a)$$ \noindent
where
$$\eqalignno{&T^{\rm Cl}_{\mu \nu}=\nabla _{\mu} \phi \nabla _{\nu} \phi -
{1\over 2}g_{\mu \nu} (\nabla \phi )^2, &({\rm 2}.30b)\cr &T_{\mu
\nu}^{\rm GB}={1\over {2\alpha V}}\biggl(\int _{S^4} dV~\phi
\biggr)^2g_{\mu \nu}-{1\over {\alpha V}}\biggl(\int _{S^4} dV~\phi
\biggr)\phi g_{\mu \nu},&({\rm 2}.30c)\cr &T_{\mu \nu}^{\rm g}=-
{\overline c}cg_{\mu \nu},&({\rm 2}.30d)\cr &T_{\mu \nu}^{\rm
b}=-{1\over 2}\alpha  b^2g_{\mu \nu}. &({\rm 2}.30e)\cr}$$
\noindent
At the quantum level $\langle T_{\mu \nu} \rangle$ is formally given
by

$$\langle T_{\mu \nu} \rangle={{\int d[\phi ]~d{\overline c}~dc~db~ T_{\mu
\nu}\exp (-S^Q)}\over {\int d[\phi ]~d{\overline c}~dc~db~\exp
(-S^Q)}}. \eqno({\rm 2}.31)$$
\noindent The calculation of the
contributions to $\langle T_{\mu \nu} \rangle$ of the ghost and
antighost fields $c$ and ${\overline c}$ and of the auxillary field
$b$ are trivial. We get (with obvious notations)
$$\eqalignno{&\langle T_{\mu
\nu}^{\rm g}  \rangle={1\over V}g_{\mu \nu},&({\rm 2}.32)\cr
&\langle T_{\mu \nu}^{\rm b}  \rangle=-{1\over 2V}g_{\mu \nu}.
&({\rm 2}.33)\cr}$$
\noindent With regard to the contribution of
$\phi$, the situation is a little more complicated. Let us remarks
that if we expand $\phi$ on the form $\phi =\sum _n a_n \phi _n$,
the coefficient $a_0$ does not appear in the expression (2.30b).
Moreover, because of (2.8), only terms of type $a_0a_n$ appear in
the expression (2.30c). As a consequence, we show that (with obvious
notations)
$$\eqalignno{&\langle T_{\mu
\nu}^{\rm Cl}  \rangle={{\prod _{n\not= 0}\int da_n~T_{\mu \nu}^{\rm
Cl}\exp(-{1\over 2}\sum _{n\not= 0}\lambda _n {a_n}^2) }\over {\prod
_{n\not= 0}\int da_n~\exp(-{1\over 2}\sum _{n\not= 0}\lambda _n
{a_n}^2) }},&({\rm 2}.34)\cr &\langle T_{\mu \nu}^{\rm GB}
\rangle={{\int da_0~ T_{\mu \nu}^{\rm GB}\exp (-{{a_0}^2 \over
{2\alpha}})}\over {\int da_0~\exp (-{{a_0}^2 \over {2\alpha}})}}.
&({\rm 2}.35)\cr}$$
\noindent The calculation of $\langle T_{\mu
\nu}^{\rm GB}  \rangle$ then gives
$$\langle T_{\mu
\nu}^{\rm GB}  \rangle=-{1\over 2V}g_{\mu \nu}.\eqno({\rm 2}.36)$$
\noindent At this level, it should be noted that the contributions
of the gauge-breaking term (2.36), of the ghost-antighost term
(2.32) and of the $b $ term (2.33) cancel. Therefore $\langle T_{\mu
\nu} \rangle$ reduces to $\langle T_{\mu \nu}^{\rm Cl} \rangle$. It
remains for us to calculate (2.34). This term needs a
regularization. By noting that $\langle T_{\mu \nu}^{\rm Cl}
\rangle$ is also obtained by the point-splitting of a quantity which
possesses a symmetric Hadamard expansion, one finds the regularized
vacuum expectation value of the stress-energy tensor [14,15].
Indeed, we have
$$\langle
T_{\mu \nu} \rangle= \langle T_{\mu \nu}^{\rm Cl}  \rangle={1\over
2}\lim _{x'\to x} ~\bigl( \nabla _{\mu} \nabla _{\nu '} - {1\over
2}g_{\mu \nu}g^{\rho \sigma '}\nabla _{\rho} \nabla _{\sigma '}
\bigr) \biggl( \sum _{n\not= 0} {{\phi _n(x) \phi _n(x')}ÊÊ\over
{\lambda _n}} + (x\leftrightarrow x') \biggr). \eqno({\rm 2}.37)$$
\noindent Now it should be noted that $\sum _{n\not= 0} {{\phi _n(x)
\phi _n(x')}ÊÊ\over {\lambda _n}}+ (x\leftrightarrow x')$ which is
given by
$$\sum _{n\not= 0} {{\phi _n(x) \phi _n(x')}ÊÊ\over {\lambda _n} } +
(x\leftrightarrow x') ={R\over 48\pi ^2}\biggl[{{1/2}\over
1-Z(x,x')}- \ln \bigl(1- Z(x,x')\bigr)  \biggr] \eqno({\rm 2}.38)$$
\noindent possesses a symmetric Hadamard expansion. We have
$$\sum _{n\not= 0} {{\phi _n(x) \phi _n(x')}ÊÊ\over {\lambda _n} }
+ (x\leftrightarrow x') ={1\over {(2\pi )^2}}\biggl[{\Delta
^{1/2}(x,x')\over \sigma(x,x')}+ V(x,x') \ln \sigma (x,x') + W(x,x')
\biggr] \eqno({\rm 2}.39)$$
\noindent where $\sigma(x,x')$ is linked
to the geodesic distance between $x$ and $x'$ by $2\sigma(x,x')=\mu
^2(x,x')$, $V(x,x')$ is a smooth geometrical function while
$W(x,x')$ is a smooth state-dependent function and $\Delta (x,x')$
is the Van Vleck determinant. (See for example [14,15] and
references therein for more details on the notation.) In the present
case, because of the maximal symmetry of (2.38), all the
coefficients of the expansion of $W(x,x')$ in powers of $\sigma
(x,x')$ are constant. Therefore, from [14,15] (see for example (3.7)
of [15]), one obviously finds that [16]
$$\langle T_{\mu \nu} \rangle _{\rm ren}={29R^2 \over 138240\pi ^2}g_{\mu \nu}.
\eqno({\rm 2}.40)$$
\noindent Of course, $\langle T_{\mu \nu}
\rangle _{\rm ren}$ is independent of the gauge parameter $\alpha $
and is maximally symmetric.

\vfill
\eject

\centerline {\bf 3 - REMARKS AND CONCLUSION}
\bigskip

           It was possible to correctly treat on $S^4$ the zero mode problem
arising in the massless minimally coupled scalar field theory and to
get a $SO(5)$-invariant quantum theory by considering it as a gauge
theory. The compactness of the background manifold has played a
crucial role. The absence of an infrared divergence in the massless
scalar field theory on $S^4$ and the $SO(5)$-invariance of the
quantum theory is easy to understand: the gauge-breaking term added
to the classical action of the theory  has allowed us to replace in
the expression of the two-point function $\langle \phi (x) \phi (x')
\rangle$ the infinite and constant term $(\phi _0)^2/\lambda _0$ by
the regular and constant one $\alpha (\phi _0)^2$.

\medskip

            On de Sitter space $dS^4$, the situation is in fact less simple. A
BRST-treatment along the lines of Section 2 is not possible because
of the infinite volume of that spacetime. Furthermore, the existence
of the infrared divergence of the two-point functions is also the
consequence of the presence of zero modes, but now the zero modes
are infinite and time-dependent. When $dS^4$ is described by a
coordinate system whose corresponding spatial sections are compact,
the complete set of mode solutions of the wave equation is discret.
It is then possible to replace the two infinite and time-dependent
zero modes by  two regular but also time-dependent zero modes. The
resulting two-point functions are then time-dependent and therefore
break the $SO(1,4)$-invariance of the spacetime [17]. Moreover, the
renormalized vacuum expectation of the stress-energy tensor is
time-dependent [18]. At the contrary, when $dS^4$ is described by a
coordinate system whose corresponding spatial sections are non
compact, the complete set of mode solutions of the wave equation is
continuous and no regularization procedure (at the level of the
two-point functions) applies without destroying the structure of the
Fock space of quantum states.

\medskip

          Because the quantum theory on $dS^4$ is not $SO(1,4)$-invariant, it
cannot be linked to the $SO(5)$-invariant quantum theory construct
on $S^4$. In particular, the Green functions of the two theories
cannot be linked by analytic continuation. Euclideanization of
spacetime is a powerful method in quantum field theory, but it must
be used with lot of care. By changing the topology of the background
manifold, it may completely change the nature of a problem and its
solution, especially when zero modes are involved.

\medskip

          A similar conclusion to ours has been obtained by Mazur and Mottola in
[22], where the problem of the conformal mode of quantum gravity is
extensively discussed. The authors question the validity of the procedure of
euclideanization in quantum gravity and avocade the necessity to
consider the lorentzian form of the path integrals.

\medskip

          Recently, many calculations have been performed involving the graviton
propagator in de Sitter space [19,20,21]. All that calculations are
done less or more explicitly on the euclidean version of de Sitter
space. They provide a graviton propagator which is
$SO(1,4)$-invariant (by analytic continuation from $S^4$ to $dS^4$)
but which presents a pathological behaviour at large distance
leading to divergences in certain physical quantities. We believe
that it could be the consequence of the treatment on $S^4$. A study
on $dS^4$ might provide a true physical graviton propagator which
could break de Sitter invariance but which is not so pathological.

\bigskip
\bigskip
\bigskip
\bigskip
\bigskip
\noindent{\bf ACKNOWLEDGEMENTS:} I wish to thank Thibault Damour for his
hospitality at IHES where this work was completed. I am grateful to Bruce
Jensen and Kenneth Nordtvedt for help with the English.

\vfill
\eject

\centerline {\bf APPENDIX}
\bigskip

            Let us consider the massive scalar field theory defined by (2.1) for
the value ${m_p}^2=-p(p+3)H^2$ with $p \in {\bf N^*}$. (It should be recalled
that in ${m_p}^2$ there is included a coupling term with the scalar curvature of
the sphere.) For such a value, the parameter $\nu $ in (2.14) is equal to
$p+{3/2}$ and therefore the two-point function (2.14) diverges. From (2.13), it
is obvious that such a divergence occurs because ${m_p}^2=-p(p+3)H^2$ is the
opposite of an eigenvalue of the laplacian. Therefore, it is a consequence of
the fact that the operator $\square + p(p+3)H^2$ possesses zero modes. As in the
massless case, this divergence in the massive theory is a gauge artifact:
indeed, for ${m_p}^2=-p(p+3)H^2$, the euclidean action (2.1) is invariant under
the $d_p$-dimensional gauge transformation
$$\phi (x) \rightarrow \phi (x) + \sum
_{i=1}^{d_p} C_i \phi _p^i (x), \eqno({\rm A}.1)$$
\noindent where
the $C_i$ are arbitrary constants; it is then necessary to quantize
it as a gauge theory.

\medskip

          Thus, as in the massless case, let us add to (2.1) the
${\bf s}$-invariant term (2.19). The difference with the massless case is the
following: in (2.18) and (2.19) the ghost field $c$, the antighost field
${\overline c}$ and the auxillary field $b$ are now space-dependent; they live in
the zero mode subspace spanned by the $d_p$ functions $\phi _p^i$. Therefore,
we will write
$$c(x)=\sum _{i=1}^{d_p} c_i \phi _p^i (x) \qquad {\overline c}(x)=\sum
_{i=1}^{d_p} {\overline c}_i \phi _p^i (x) \qquad b(x)=\sum
_{i=1}^{d_p} b_i \phi _p^i (x). \eqno({\rm A}.2)$$
\noindent Then,
by expanding the scalar field $\phi$ as
$$\phi (x)=\sum _n \sum
_{i=1}^{d_n} a_{n,i}\phi _n^i (x),\eqno({\rm A}.3)$$
\noindent and
by using the normalization relation (2.5) and by performing the
shifts $b_i \rightarrow b_i + {1\over \alpha}(a_{p,i})$, we obtain
for the quantum action
$$S^Q(\phi,c,{\overline c},b)={1\over 2}\sum _{n\not= p} \sum
_{i=1}^{d_n} (\lambda _n +{m_p}^2){(a_{n,i})}^2 + {1\over {2\alpha
}}\sum _{i=1}^{d_p} {(a_{p,i})}^2 - \sum _{i=1}^{d_p}{\overline
c}_ic_i - {1\over 2}\alpha \sum _{i=1}^{d_p} {b_i}^2.\eqno({\rm
A}.4)$$
\noindent As in the massless case, it is necessary in the
calculations to rotate the integration contours for the $b_i$, $c_i$
and ${\overline c}_i$, or equivalently to integrate over the real
values of these variables but with the quantum action
$$S^Q(\phi,c,{\overline c},b)={1\over 2}\sum _{n\not= p} \sum
_{i=1}^{d_n} (\lambda _n +{m_p}^2){(a_{n,i})}^2 + {1\over {2\alpha }}\sum
_{i=1}^{d_p}
{(a_{p,i})}^2 + \sum
_{i=1}^{d_p}{\overline c}_ic_i + {1\over 2}\alpha \sum
_{i=1}^{d_p} {b_i}^2.\eqno({\rm
A}.5)$$

\medskip

        The true two-point function $G(x,x';{m_p}^2)=\langle \phi (x) \phi
(x') \rangle$ can then be easily obtained. By inserting (A.3) and (A.5) in
(2.26), one finds
$$G(x,x';{m_p}^2)= \sum _{n\not= p} \sum
_{i=1}^{d_n}{{\phi _n^i(x) \phi _n^i(x')}ÊÊ\over {\lambda
_n}+{m_p}^2 } + \alpha  \sum _{i=1}^{d_p}{\phi _p^i(x) \phi
_p^i(x')}. \eqno({\rm A}.6)$$ \noindent All the sums $\sum
_{i=1}^{d_n}{\phi _n^i(x) \phi _n^i(x')}$ are $SO(5)$-invariant [the
${(\phi _n^i)}_{i=1,...,d_n}$ form a basis of the $d_n$-dimensional
representation of $SO(5)$] and are given by [20]
$$\sum _{i=1}^{d_n}{\phi _n^i(x) \phi
_n^i(x')} = {6d_n \over 16\pi ^2}H^4 F \bigl[ -n,n+3;2;1-Z(x,x')
\bigr]. \eqno({\rm A}.7)$$ \noindent (Here the expansion of the
hypergeometric function $F$ terminates and in fact that sum reduces
to a Gegengauer polynomial.) Therefore, the two-point function
$G(x,x';{m_p}^2)$ is also  $SO(5)$-invariant and the quantum theory
possesses this invariance. Moreover, as in the massless case, and
because of the choice of the gauge-fixing term, the quantum theory
is independent of the gauge parameter $\alpha$.

\medskip

           To conclude this appendix let us note i) that the massless case
can be considered as the particular case $p=0$ in the previous
calculations, ii) that on $dS^4$ the divergences appearing in the
quantum theories for the mass values ${m_p}^2=-p(p+3)H^2$ with $p
\in {\bf N^*}$ are not gauge artifacts: they are real and correspond
to the impossibility to construct $SO(1,4)$-invariant theories. That
last point has been study in [23] in the case $p=1$ which described
the scalar part of the metric fluctuation.

\vfill
\eject

\centerline{\bf REFERENCES}
\bigskip
\bigskip
\parindent=0.0cm

\smallskip
[1] G.W. Gibbons, in {\it General Relativity, An Einstein Centenary Survey},
ed.~by S.W. Hawking and W. Israel (Cambridge University Press, Cambridge, 1979).

\smallskip
[2] S.W. Hawking, in {\it General Relativity, An Einstein Centenary Survey},
ed.~by S.W. Hawking and W. Israel (Cambridge University Press, Cambridge, 1979).

\smallskip
[3] R. Wald, Commun. Math. Phys. {\bf 70}, 221 (1979).

\smallskip
[4] A. Meister, J. Math. Phys. {\bf 30}, 2930 (1989).

\smallskip
[5] S.W. Hawking, in {\it Relativity, Groups and Topology II}, ed.~by B.S.
DeWitt and R. Stora (North Holland, Amsterdam, 1984).

\smallskip
[6] B. Allen,  Phys. Rev. {\bf D32}, 3136 (1985).

\smallskip
[7] G.W. Gibbons and M.J. Perry, Nucl. Phys. {\bf B146}, 90 (1978).

\smallskip
[8] B. Allen and T. Jacobson, Commun. Math. Phys. {\bf 103}, 669 (1986).

\smallskip
[9] M. Abramowitz and I.A. Stegun, {\it Handbook of Mathematical Functions}
, (Dover, N.Y., 1965).

\smallskip
[10] L. Baulieu and J. Thierry-Mieg, Nucl. Phys. {\bf B197}, 477 (1982).

\smallskip
[11] L. Baulieu, Phys. Rep. {\bf 129}, 1 (1985).

\smallskip
[12] L. Baulieu and M. Bellon, Phys. Lett. {\bf B202}, 67 (1988).

\smallskip
[13] The same difficulty exists for electromagnetism (a theory which has no
infrared divergence problem). Indeed, for electromagnetism, the classical
action $S={1\over 4}\int _{S^4} dV~ F^2$ is invariant under the gauge
transformation $A_{\mu} \rightarrow A_{\mu} + V_{;\mu}$. The BRST-quantization
of the theory is realized by introducing a ghost field $c$, an antighost field
${\overline c}$, an auxillary field $b$ and a fermionic operator ${\bf s}$.
(Here $c$, ${\overline c}$ and $b$ are non-constant.) The operator ${\bf s}$ is
such that ${\bf s}A_{\mu} =c_{;\mu}$, ${\bf s}c =0$, ${\bf s}{\overline c}=b$
and ${\bf s}b =0$. The gauge-fixing term is now $S^{GF}={\bf s}\int _{S^4}
dV~\bigl({\overline c}A_{\mu}{}^{;\mu} - {1\over 2}{\overline c}\alpha b \bigr)$
and by shifting $b$ we get for the quantum action $S^Q(A_{\mu},c,{\overline
c},b)={1\over 4}\int _{S^4} dV~ F^2 + {1\over {2\alpha
}}\int _{S^4} dV~(A_{\mu}{}^{;\mu})^2 -\int _{S^4}dV~{\overline c}\square c -
{1\over 2}\alpha \int _{S^4} dV~b^2$. The $b$ term in the action is
still problematic. If we decompose $b$ on the $\phi _n$ as $b=\sum _n b _n \phi
_n$, we get $-{1\over 2}\alpha \int _{S^4} dV~b^2 = -{1\over
2}\alpha \sum _n {b_n}^2$. Therefore, in order to obtain well-defined path
integrals, it is necessary to integrate all the $b_n$ from $-i\infty$ to
$+i\infty$. On the contrary, the ghost and antighost fields are now propagating
fields and no change of sign of the corresponding action is necessary.

\smallskip
[14] M.R. Brown and A.C. Ottewill, Phys. Rev. {\bf D34}, 1776 (1986).

\smallskip
[15] D. Bernard and A. Folacci, Phys. Rev. {\bf D34}, 2286 (1986).
\parindent=0.0 truecm

\smallskip
[16] In the massive case, the euclidean vacuum expectation value of the
stress-energy tensor has been calculated by several authors. (See for example
[T.S. Bunch and P.C.W. Davies, Proc. R. Soc. {\bf A360}, 117 (1978)].) In the
massless limit, their result reduces to $\langle T_{\mu \nu} \rangle _{\rm
ren}=-(61R^2/138240\pi ^2)g_{\mu \nu}$. It differs from (2.40), but that is not
really surprising: in the massive case, $\langle T_{\mu \nu} \rangle _{\rm
ren}$ is obtained by point-splitting and regularization from $G(x,x';m^2)$; by
looking at (2.15) and by using (3.7) of [15], we understand that the massless
limit of the Bunch-Davies result and the result (2.40) exactly differ by
$(m^2g_{\mu \nu}/4)\times ({R^2/384\pi ^2m^2})$ which is equal
to $(90R^2/138240\pi ^2)g_{\mu \nu}$.

\smallskip
[17] B. Allen and A. Folacci,  Phys. Rev. {\bf D35}, 3771 (1987).

\smallskip
[18] A. Folacci, J. Math. Phys. {\bf 32}, 2828 (1991) and erratum in
J. Math. Phys. {\bf 33}, 1932 (1992).

\smallskip
[19] B. Allen, Phys. Rev. {\bf D34}, 3670 (1986); B. Allen and M. Turyn, Nucl.
Phys. {\bf B292}, 813 (1987).

\smallskip
[20] E.G. Floratos, J. Iliopoulos and T.N. Tomaras, Phys. Lett. {\bf B197}, 373
(1987).

\smallskip
[21] I. Antoniadis and E. Mottola, J. Math. Phys. {\bf 32}, 1037 (1991).

\smallskip
[22] P.O. Mazur and E. Mottola, Nucl. Phys. {\bf B341}, 187 (1990).

\smallskip
[23] P.O. Mazur and E. Mottola, Nucl. Phys. {\bf B278}, 694 (1986).

\vfill
\eject

\bye